\documentclass[prx,a4paper,10pt,twocolumn,superscriptaddress,aps,nolongbibliography]{revtex4-2}

\usepackage{graphicx,color}
\usepackage{dcolumn}
\usepackage{bm}
\usepackage{amsmath,amssymb,mathrsfs}
\usepackage{physics}
\usepackage{url}
\usepackage[colorlinks=true,
            citecolor=blue]{hyperref}

\newcommand{\R}{\mathbb{R}}

\newcommand{\id}{\mathbb{I}}
\newcommand{\pseq}{p^{\mathrm{seq}}}

\newcommand{\pLPT}{p^{\mathrm{LPT}}}
\newcommand{\pPM}{p^{\mathrm{PM}}}
\newcommand{\pT}{p^{\mathrm{T}}}
\newcommand{\pI}{p^{\mathrm{I}}}
\newcommand{\pC}{p^{\mathrm{C}}}
\newcommand{\pD}{p^{\mathrm{D}}}

\newcommand{\set}[1]{\mathsf{#1}}

\newcommand{\Span}{{\mathsf{Span}}}

\def\>{\rangle}
\def\<{\langle}

\newcommand{\map}[1]{\mathcal{#1}}

\newcommand{\St}{{\mathsf{St}}}
\newcommand{\Eff}{{\mathsf{Eff}}}

\newcommand{\Transf}{{\mathsf{Transf}}}

\newtheorem{theo}{Theorem}

\newtheorem{lemma}{Lemma}
\newtheorem{prop}{Proposition}
\newtheorem{cor}{Corollary}
\newtheorem{defi}{Definition}

\newtheorem{proper}{Property}

\def\Proof{{\bf Proof.~}}
\def\qed{$\blacksquare$ \medskip}

\newcommand{\indep}{\perp \!\!\! \perp}

\begin{document}
    \title{
    Local causal consistency as a foundational principle for indefinite causal order
    }

    \author{Zixuan Liu}
    \email{zixuanliu.cs@gmail.com}
    \affiliation{QuIC, Ecole Polytechnique de Bruxelles, C.P. 165, Universit\'e Libre de Bruxelles, 1050 Brussels, Belgium}
    \author{Ognyan Oreshkov}
    \email{ognyan.oreshkov@ulb.be}
    \affiliation{QuIC, Ecole Polytechnique de Bruxelles, C.P. 165, Universit\'e Libre de Bruxelles, 1050 Brussels, Belgium}

\begin{abstract}

    Quantum processes with indefinite causal order challenge the description of physical experiments as globally ordered networks while retaining causality within each local laboratory. Whether and how this local causal requirement constrains the information flow in such quantum processes remains unclear. Here we formalize the principle of local causal consistency through experiments in which each laboratory performs two sequential operations and may use the earlier outcome to choose the later operation. The principle is equivalent to a set of parity-erasure conditions: the joint earlier outcomes for any set of laboratories reveal no information about the parity of the corresponding independent, uniformly random binary settings. We show that these conditions precisely characterize, via local process tomography, the multipartite channels representing valid quantum processes; characterize classical process-matrix correlations through local readout of outcome tables; and, together with the gentle-measurement property, constrain general quantum correlations with indefinite causal order. Moreover, these results extend naturally to a vast class of theories beyond quantum theory. Our findings provide an operational account of how causality within local laboratories shapes indefinite causal structure and the correlations it supports, revealing a fundamental constraint on information exchange through indefinite causal structure.
\end{abstract}

\maketitle

\section{Introduction}
The notion of causality, which prohibits signaling from the future to the past, is central to our understanding of the physical world. In quantum physics, processes are conventionally formulated as causal networks in which the connections between operations mediate causal influences, and all events are compatible with a well-defined causal order. 

Foundational studies of dynamical causal structure have shown that quantum theory accommodates processes with indefinite causal order (ICO) \cite{hardy2007towards,chiribella2013quantum,oreshkov2012quantum,oreshkov2016causal,wechs2019definition}, which can offer advantages in a variety of information-processing tasks (see, e.g., Refs. \cite{chiribella2012perfect,araujo2014computational,guerin2016exponential,ebler2018enhanced,zhao2020quantum,felce2020quantum,gao2023measuring,zhu2023charging}). 
Such processes are described by the process matrix formalism, and can violate Bell-like inequalities for causal order \cite{oreshkov2012quantum,branciard2015simplest,oreshkov2016causal,abbott2016multipartite,van2023device}, known as causal inequalities. They are expected to arise at the interface of quantum mechanics and general relativity \cite{zych2019bell,moller2021quantum}
and have also been shown to exist on time-delocalized quantum subsystems in standard quantum mechanics 
\cite{oreshkov2019time,wechs2023existence}.
In the process matrix formalism, operations within local laboratories are described by standard quantum theory, whereas the external environment connecting the laboratories is characterized by a process matrix. This matrix represents a higher-order quantum process, also known as a quantum supermap \cite{chiribella2013quantum}, and is equivalent to a particular type of channel connected to the local operations in a cyclic network. 

The existence of ICO highlights a profound tension. On the one hand, it implies that a complete causal description of experiments may not be necessary in physics. On the other hand, it emphasizes the validity of causality within local laboratories. Operationally, sequential events within a local laboratory must obey no backward signaling \cite{chiribella2010probabilistic,brukner2014talk,guryanova2019exploring}: the choice of a later operation cannot influence the outcome of an earlier one, which rules out unrestricted post-selected quantum processes \cite{oreshkov2016operational,silva2017connecting,guryanova2019exploring}.

These observations raise fundamental questions about the information-theoretic role of causality in higher-order quantum processes, which can be examined from two perspectives. The first concerns the structure of higher-order quantum processes, which has been investigated using algebraic, type-theoretic, and combinatorial methods \cite{araujo2015witnessing,bisio2019theoretical,hoffreumon2026projective,jenvcova2026structure}. Viewed as a channel, a quantum process can be specified by its input-output relations. An intriguing question is how locally valid causality translates into conditions for indefinite causal structure, and in turn into global conditions on the input-output relations of quantum processes.

The second concerns the fundamental limits of correlations arising from higher-order quantum processes. Although the process-matrix framework permits violations of causal inequalities, recent works \cite{kunjwal2023nonclassicality,liu2025tsirelson} show that quantum theory imposes nontrivial restrictions on the attainable violations, indicating that quantum correlations cannot be noncausal to an arbitrary degree. In circuit-based operational probabilistic theories, causality admits a precise information-theoretic formulation \cite{chiribella2010probabilistic}, and in Bell scenarios, quantum limits such as Tsirelson's bound \cite{cirel1980quantum} have been shown to follow from informational principles \cite{van1999nonlocality,brassard2006limit,brunner2009nonlocality,linden2007quantum,pawlowski2009information,navascues2010glance,fritz2013local,wakakuwa2021gentle}. In contrast, in the process-matrix framework, it remains unclear whether, and through what information-theoretic mechanism, causality within local laboratories constrains quantum correlations across them.

Here we address both questions using a paradigm of locally sequential experiments, in which every laboratory performs two sequential quantum instruments. We propose to formalize local causal consistency (LCC) as a stronger version of no backward signaling, requiring that the experiment remain valid under arbitrary local feed-forward, whereby the outcome of the first instrument determines the choice of the second.

We first show that LCC is equivalent to the validity condition of classical process matrices. This condition, in turn, is equivalent to a set of \emph{parity-erasure conditions}: for independent, uniformly random binary settings of the later instruments, the earlier outcomes associated with any subset of laboratories contain no information about the parity of the corresponding later settings. Through local process tomography, we prove that a multipartite channel corresponds to a valid process matrix if and only if its tomographic statistics satisfy the parity-erasure conditions. This result gives the linear constraints defining process matrices an operational interpretation and connects process matrices to a parity-erasing variant of random access codes.

We then investigate the implications of LCC for the correlations observed between laboratories. Restricting to classical processes, we prove that a correlation is realizable if and only if it is a local readout of an outcome-table distribution satisfying LCC. In the general case with quantum processes, we relate correlations to locally sequential experiments with gentle measurements, through which LCC yields quantitative bounds on two-way signaling and excludes a neighborhood of every deterministic correlation without a classical realization.

These conclusions extend beyond quantum theory: the parity-erasure characterization of higher-order processes holds in any causal operational probabilistic theory satisfying the properties called local tomography and one-way-signaling decomposability, while the correlation bounds can be derived from LCC and an appropriate gentle-measurement property.

Taken together, our results clarify how causality within local laboratories shapes both the global structure of higher-order quantum processes and the correlations they can generate, and thus establish LCC as an information-theoretic basis for quantum theory with ICO.

\section{Results}
\subsection*{Locally sequential experiments}
Consider $N$ separate experimenters, each of whom can freely choose an operation within their own laboratory. The operation of the $i$-th experimenter is described by a quantum instrument, represented as a collection of completely positive maps $\big( \map M_{m_i}^{(i)} \big)$ from input quantum system $A_i$ to output quantum system $B_i$, with an associated outcome $m_i$. In the process matrix formalism \cite{oreshkov2012quantum}, the environment outside the laboratories is characterized by a process matrix $S$ that acts on the composite system $A_1 \ldots A_N B_1 \ldots B_N$. The process matrix describes a quantum supermap that assigns joint probabilities $\Pr_S\!\big[\mathcal{M}_{m_1}^{(1)}, \ldots, \mathcal{M}_{m_N}^{(N)}\big]$ for any choice of local operations. Viewed as a cyclic network, a quantum supermap is a specific type of channel that maps the joint output $B_1 \ldots B_N$ of the local laboratories to their joint input $A_1 \ldots A_N$.

\begin{figure}[htbp]
    \centering
    \includegraphics{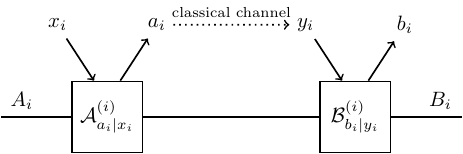}
    \caption{\emph{Local causal consistency.} Within each laboratory, the outcome of an earlier test may determine the setting of a subsequent test, whereas the later setting cannot influence the earlier outcome.}
    \label{fig:causality}
\end{figure}

Here we focus on local operations comprising two sequential quantum instruments, ensuring a well-defined order within each laboratory. As illustrated in Fig.~\ref{fig:causality}, the $i$-th experimenter first performs an instrument $\big( \mathcal{A}^{(i)}_{a_i \mid x_i} \big)$ depending on a setting $x_i$, followed by an instrument $\big( \mathcal{B}^{(i)}_{b_i \mid y_i} \big)$ depending on a setting $y_i$. Let $\bm{x} = (x_1,\ldots,x_N)$ denote the list of settings for the first instruments, and similarly define $\bm{y}$, $\bm{a}$, and $\bm{b}$ for the second settings and the outcomes of both instruments. Throughout this paper, all settings and outcomes are drawn from finite alphabets. The statistics of a locally sequential experiment are then described by a conditional probability distribution \mbox{$\pseq(\bm{a},\bm{b} \mid \bm{x},\bm{y})$}.

\subsection*{Local causal consistency}
A necessary consequence of causality within the $i$-th laboratory is no signaling from the later instrument to the earlier one: the setting $y_i$ cannot influence the outcome $a_i$. Formally, the marginal distribution \mbox{$\pseq(a_i\mid\bm x,\bm y)$} must be independent of $y_i$. This condition constrains only the signaling relations within individual laboratories and does not capture possible correlations among the earlier outcomes of different laboratories.

A stronger operational requirement follows from closure under conditioned tests \cite{chiribella2010probabilistic}: an earlier outcome
may be used to choose a later operation. Accordingly, in each laboratory the outcome $a_i$ may determine the subsequent setting $y_i$ through an arbitrary classical channel \mbox{$\kappa_i(y_i\mid a_i)$}, as illustrated in Fig.~\ref{fig:causality}. After inserting these local feed-forward channels, the resulting statistics are
\begin{equation}
    \label{eq:classical}
    p(\bm a,\bm b,\bm y\mid\bm x)
    =
    \pseq(\bm a,\bm b\mid\bm x,\bm y)
    \prod_{i=1}^{N}\kappa_i(y_i\mid a_i)\, .
\end{equation}
We say that the statistics \mbox{$\pseq(\bm{a},\bm{b} \mid \bm{x},\bm{y})$} are compatible with LCC when Eq.~\eqref{eq:classical} defines a valid conditional probability distribution for every collection of local channels $\kappa_i$. Equivalently, the open statistics $\pseq$ define a higher-order transformation that maps a list of local classical channels to a probability distribution, as in the classical process-matrix framework \cite{oreshkov2012quantum,baumeler2016space}.

All factors in Eq.~\eqref{eq:classical} are nonnegative, so normalization is the only nontrivial validity requirement. Summing over $\bm b$ shows that this requirement depends only on the marginal distribution \mbox{$\pseq(\bm a\mid\bm x,\bm y)$}; we therefore suppress $\bm b$ below. 
In other words, LCC can be understood as a restriction on the marginal distribution \mbox{$\pseq(\bm a\mid\bm x,\bm y)$} so that it defines a classical process matrix for every $\bm x$. Mathematically, the restriction can be formulated as the linear constraints characterizing process matrices \cite{araujo2015witnessing,chiribella2016optimal}. We next express these constraints directly in terms of the observable probabilities.

\subsection*{Parity-erasure conditions}
We first consider two parties with binary later settings \mbox{$y_1,y_2\in\{0,1\}$}. In this case, the validity of Eq.~\eqref{eq:classical} is equivalent to the following linear constraints, which hold for all settings $x_1,x_2$ and all relevant outcomes $a_1,a_2$:
\[
    \begin{split}
        &\sum_{y_1,y_2\in\{0,1\}}(-1)^{y_1}
        \pseq(a_1\mid x_1,x_2,y_1,y_2)=0\, ,\\
        &\sum_{y_1,y_2\in\{0,1\}}(-1)^{y_2}
        \pseq(a_2\mid x_1,x_2,y_1,y_2)=0\, ,\\
        &\sum_{y_1,y_2\in\{0,1\}}(-1)^{y_1\oplus y_2}
        \pseq(a_1,a_2\mid x_1,x_2,y_1,y_2)=0\, .
    \end{split}
\]

When $y_1$ and $y_2$ are independent and uniformly distributed random bits, these constraints have a direct information-theoretic interpretation as the independence conditions of random variables:
\begin{align*}
    &(a_1\indep y_1)\mid(x_1,x_2)\, ,\\
    &(a_2\indep y_2)\mid(x_1,x_2)\, ,\\
    &((a_1,a_2)\indep y_1\oplus y_2)\mid(x_1,x_2)\, .
\end{align*}
Thus, each earlier outcome contains no information about the corresponding later setting, while the joint earlier outcomes contain no information about the parity of the two later settings. For $N$ parties with independent, uniformly distributed binary settings, the same interpretation applies to every nonempty subset $I\subseteq\{1,\ldots,N\}$:
\[
    \bigg(\bm a_I \indep \bigoplus_{i\in I} y_i \bigg) \mid \bm x\, ,
\]
where $\bm a_I=(a_i)_{i\in I}$. In other words, the earlier outcomes of any subset of laboratories reveal no information about the parity of the corresponding later settings. We therefore refer to these relations as the \emph{parity-erasure conditions}.

These conditions extend to arbitrary setting alphabets by selecting two, not necessarily distinct, settings for each party. Specifically, let $w_i\in\{0,1\}$ be a binary variable and let $f_i$ encode it into a setting, $y_i=f_i(w_i)$. We use the shorthand $f(\bm w)=(f_1(w_1),\ldots,f_N(w_N))$. The general result is as follows (see the proof in Methods).
\begin{theo}
\label{theo:parity}
Eq.~\eqref{eq:classical} is valid for every collection of local classical channels if and only if, for every nonempty subset $I\subseteq\{1,\ldots,N\}$ and every collection of encoding functions $f_i$, the following constraints are satisfied for all $\bm x$ and $\bm a_I$:
\begin{equation}
    \label{eq:pe}
    \sum_{\bm w\in\{0,1\}^N}
    (-1)^{\sum_{i\in I}w_i}
    \pseq\big(\bm a_I\mid\bm x,f(\bm w)\big)=0\, .
\end{equation}
\end{theo}

\subsection*{Tomographic characterization of process matrices}
The parity-erasure conditions derived above provide a
key insight into the structure of process matrices: they establish a sharp information-theoretic boundary between them and all other quantum channels. This is demonstrated through a local process tomography procedure, defined as follows:
\begin{defi}
    \label{defi:lpt}
    Let $\map T$ be a quantum channel from system $B_1\ldots B_N$ to system $A_1\ldots A_N$. A \emph{local process tomography} (LPT) of $\map T$ consists, for each party $i$, of a tomographically complete family of input states $\{\rho^{B_i}_{y_i}\}_{y_i}$ and a tomographically complete family of measurements $\{e^{A_i}_{a_i\mid x_i}\}_{a_i,x_i}$. These local probes represent the channel by the conditional probability distribution
    \[
        \pLPT(\bm a\mid\bm x,\bm y)
        :=\bigotimes_{i=1}^N e^{A_i}_{a_i\mid x_i}
        \left[
        \map T\left(\bigotimes_{i=1}^N\rho^{B_i}_{y_i}\right)
        \right] \, .
    \]
\end{defi}

Because the local preparations and measurement effects are tomographically complete, $\pLPT$ uniquely determines the channel $\map T$. Moreover, in the context of process-matrix experiments, the LPT scheme has the temporal structure of the locally sequential paradigm: within each laboratory, the measurement on $A_i$ is performed first, and the preparation on $B_i$ is performed second and has a trivial outcome. 

In the LPT scheme, we show that, within the set of all multipartite channels, the subset corresponding to valid process matrices emerges when parity-erasure conditions are imposed:
\begin{theo}
    \label{theo:characterization}
    Let \mbox{$\pLPT(\bm a\mid\bm x,\bm y)$} be an LPT representation of an $N$-partite quantum channel $\map T$. The following statements are equivalent:
    \begin{enumerate}
        \item $\map T$ represents an $N$-party process matrix;
        \item the LPT statistics $\pLPT$ are compatible with LCC;
        \item the LPT statistics $\pLPT$ satisfy the parity-erasure conditions in Eq.~\eqref{eq:pe}.
    \end{enumerate}
\end{theo}

Theorem~\ref{theo:characterization} identifies parity erasure as a necessary and sufficient information-theoretic condition for a multipartite quantum channel to represent a process matrix. It also suggests an interpretation in terms of a distributed encoding task, related to random access codes \cite{wiesner1983conjugate,ambainis1999dense,ambainis2002dense} and, in particular, parity-oblivious random access codes \cite{spekkens2009preparation}. Here, a binary string $\bm y$ is encoded into a multipartite state, with the requirement that specified groups of output systems reveal no information about the parity of the corresponding input bits. The following corollary states this property directly, serving as the channel-level counterpart of the parity-erasure conditions.
\begin{cor}
    \label{theo:rac}
    Let $\map T$ be a quantum channel from system $B_1\ldots B_N$ to system $A_1\ldots A_N$. For each party $i$, choose two input states $\rho^{B_i}_{0}$ and $\rho^{B_i}_{1}$, and define
    \[
        \Psi_{\bm y}
        :=\map T\left(\bigotimes_{i=1}^N\rho^{B_i}_{y_i}\right),
        \qquad \bm y\in\{0,1\}^N\, .
    \]
    Then $\map T$ represents an $N$-party process matrix if and only if, for every choice of these local input states and every nonempty subset $I\subseteq\{1,\ldots,N\}$,
    \[
        \Tr_{A_{I^c}}
        \left[
        \sum_{\bm y\in\{0,1\}^N}
        (-1)^{\sum_{i\in I}y_i}\Psi_{\bm y}
        \right]=0\, ,
    \]
    where $A_{I^c}:=\bigotimes_{i\notin I}A_i$.
\end{cor}

Although stated above in quantum terms, Theorem~\ref{theo:characterization} and Corollary~\ref{theo:rac} do not require the formalism of quantum theory. As shown in Methods, both results hold in any causal probabilistic theory satisfying local tomography and one-way-signaling decomposability (OWSD), as defined below.

\begin{proper}
    [Local tomography]
    Every multipartite state is uniquely determined by the statistics of local measurements on its component systems.
\end{proper}

\begin{figure}[htbp]
    \centering
    \includegraphics{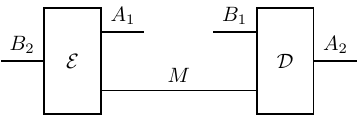}
    \caption{\emph{The one-way-signaling circuit.} Under the assumption of one-way-signaling decomposability, every channel satisfying the no-influence condition $B_1 \not\to A_1$ can be implemented by a circuit with a slot in which an arbitrary operation from system $A_1$ to system $B_1$ can be inserted.}
    \label{fig:causaldecom}
\end{figure}

\begin{proper} [OWSD]
    Every bipartite channel from system $B_1 B_2$ to system $A_1 A_2$ that satisfies the no-influence condition \mbox{$B_1 \not\to A_1$} (that is, the marginal output of system $A_1$ does not depend on the input on system $B_1$) can be implemented by connecting a channel $\map E$ from $B_2$ to $A_1M$ and a channel $\map D$ from $B_1M$ to $A_2$, through a memory system $M$, as visualized in Fig.~\ref{fig:causaldecom}.
\end{proper}

Local tomography allows transformations of states to be characterized using local
preparations and measurements \cite{chiribella2010probabilistic}, and thus
enables local process tomography. For classical channels, OWSD follows from the chain rule of probability theory. It also holds for quantum channels \cite{eggeling2002semicausal} and, more generally, in causal operational probabilistic theories satisfying local tomography and purification \cite{chiribella2010probabilistic}. This guarantees that any channel in which a given input subsystem cannot influence a given output subsystem admits an explicit circuit realization which we call a one-way-signaling circuit, where an arbitrary transformation acting between
the two subsystems can be probed. This one-way-signaling circuit, illustrated in Fig.~\ref{fig:causaldecom}, is the key ingredient in the recursive proof of Theorem~\ref{theo:characterization}.

\subsection*{Classical process-matrix correlations}

We now turn to the implications of LCC for the correlations observed between laboratories. In a general process-matrix experiment, party $i$ chooses a setting $z_i$, applies an instrument $\big(\map M^{(i)}_{m_i\mid z_i}\big)$, and records an outcome $m_i$. The joint outcome distribution is called a \emph{process-matrix correlation}:
\begin{align}
    \label{eq:general}
    \pPM(\bm m\mid\bm z)
    &=
    \Pr_S\!\big[
        \map M^{(1)}_{m_1\mid z_1},
        \ldots,
        \map M^{(N)}_{m_N\mid z_N}
    \big] \nonumber \\
    &= \Tr[S^T \bigotimes_{i=1}^N \operatorname{Choi}(\map M^{(i)}_{m_i\mid z_i})]
    \, .
\end{align}
Here $\operatorname{Choi}(\map M)$ denotes the Choi operator of $\map M$. As in Bell scenarios, we make no assumptions about the system dimensions or the internal structure of local operations.

Process-matrix correlations can violate causal inequalities \cite{oreshkov2012quantum}, and such violations also occur in the classical process-matrix framework with three or more parties \cite{baumeler2016space}. Quantum theory, nevertheless, imposes nontrivial upper bounds on the attainable violations. For example, Ref.~\cite{liu2025tsirelson} established that, for independent, uniformly random binary settings, the guess-your-neighbor's-input (GYNI) success probability satisfies \mbox{$\Pr_{\pPM}(m_1=z_2,m_2=z_1)\leq0.7592$}. 

These bounds motivate us to ask what restrictions on process-matrix correlations follow from the principle of LCC. This question requires connecting general experiments to locally sequential ones where LCC can be imposed. Here we consider locally sequential settings in which every local laboratory applies an instrument that records a table of outcomes, followed by an instrument chosen according to the setting $z_i$. Let $\mathcal Z_i$ be party $i$'s setting alphabet and let $\bm\alpha_i=(\alpha_{i,j})_{j\in\mathcal Z_i}$ be a table of outcomes recorded in the $i$-th laboratory before $z_i$ is chosen, with one entry for each possible setting. Denote the joint table distribution by \mbox{$\pT(\bm\alpha\mid\bm z)$}, where $\bm\alpha=(\bm\alpha_1,\ldots,\bm\alpha_N)$. With the tables as earlier outcomes and $\bm z$ as later settings, $\pT$ must satisfy LCC, or equivalently the parity-erasure conditions. We also consider the local readout of the selected entries, $m_i=\alpha_{i,z_i}$, which gives a correlation of the form in Eq.~\eqref{eq:general}:
\begin{equation}
    \label{eq:classical-embedding}
    p(\bm m\mid\bm z)
    =
    \sum_{\bm\alpha}
    \left[
        \prod_{i=1}^{N}\delta_{m_i,\alpha_{i,z_i}}
    \right]
    \pT(\bm\alpha\mid\bm z)\, .
\end{equation}

We show that LCC characterizes classical process-matrix correlations through Eq.~\eqref{eq:classical-embedding}.
A classical instrument in laboratory $i$ can be represented as a conditional distribution \mbox{$\mu_i(s_i',m_i\mid s_i,z_i)$}, where $s_i$ and $s_i'$ are the input and output classical states. Since $s_i$ can be copied without disturbance, the instruments for all settings can be applied to separate copies before $z_i$ is chosen. Their outcomes form $\bm\alpha_i$, and their output states are retained. Once $z_i$ is chosen, a later instrument sends the corresponding output state to the process. Reading out $\alpha_{i,z_i}$ then reproduces the original classical instrument \mbox{$\mu_i(s_i',m_i\mid s_i,z_i)$}. Hence, any classical process-matrix correlation can be realized by first producing a table distribution and then reading out the outcome locally.
Conversely, any table distribution satisfying LCC defines a valid classical process that maps the outgoing settings $\bm z$ to the incoming tables $\bm\alpha$. A local readout is implemented by classical operations in which each party $i$ reports $m_i=\alpha_{i,z_i}$ and sends $z_i$ to the process. Hence, Eq.~\eqref{eq:classical-embedding} represents a classical realization.

In other words, a correlation is classically realizable if and only if it is a local readout of an LCC-compatible table distribution.

The analysis for classical correlations cannot be directly extended to general correlations, as the copying step is unavailable for general quantum inputs. Ref.~\cite{kunjwal2023nonclassicality} nevertheless proved that a deterministic correlation admits a quantum process-matrix realization if and only if it admits a classical one. In their proof, a projective instrument that leaves the input states arising in the experiment undisturbed is inserted in each laboratory before the local setting is chosen. This produces a deterministic LCC-compatible table distribution, and its local readout recovers the deterministic correlation.

Inspired by the deterministic-case proof, we next insert gentle measurements into general process-matrix experiments to construct LCC-compatible outcome tables. Their local readout approximates the original correlations with an error controlled by the measurement disturbance, allowing us to derive quantitative correlation bounds.

\subsection*{Correlation bounds from a gentle-measurement principle}

To illustrate the idea, suppose that process-matrix correlations could achieve a GYNI success probability arbitrarily close to one. Each laboratory's input would then contain nearly perfect information about the other party's setting. Before choosing its own setting, each party could record a guess using a gentle measurement associated with a fixed setting and then perform the original setting-dependent operation. As GYNI success approached one, the disturbance from these insertions would vanish, and the joint earlier records would reveal the parity of the later settings with high probability, contradicting parity erasure and hence LCC. The GYNI success probability must therefore be bounded strictly below one.

To quantify this argument, we use the gentle-measurement lemma \cite{winter1999coding,winter1999codingt,winter2001capacity,ogawa2007making}, which links outcome predictability to state disturbance. Ref.~\cite{wakakuwa2021gentle} developed this link into an information-theoretic principle that rules out Bell correlations attaining the algebraic maximum of CHSH and, through information causality, recovers Tsirelson's bound.

Here we adopt a gentle-measurement property for the non-selective operation induced by the measurement, rather than for the operation associated with a particular outcome:

\begin{proper}[Gentle measurement]
    \label{prop:gentle-measurement}
    Every measurement $(e_a)$ on a system $A$ can be associated with a gentle implementation, that is, an instrument $(\map G_a)$ from $A$ to itself. On every state $\rho^A$, this instrument produces the measurement outcome distribution
    \[
        p_e(a\mid\rho^A) = e_a(\rho^A) = \Tr[\map G_a(\rho^A)] \, .
    \]
    Moreover, for every reference system $R$ and joint state $\rho^{AR}$, the non-selective action of the instrument satisfies
    \begin{equation}
        \label{eq:gentle-measurement}
        \frac12\left\|\left[\left(\sum_a\map G_a\right)\otimes\map I^R\right](\rho^{AR})-\rho^{AR}\right\|_1
        \leq\eta(\varepsilon)\, ,
    \end{equation}
    where
    \[
    \varepsilon=1-\max_a p_e(a\mid\rho^A) \, , \quad
    \rho^A = \Tr_{R} \rho^{AR}\, ,
    \]
    and $\eta: [0,1]\to[0,1]$ is a fixed function, continuous and nondecreasing, with $\eta(0)=0$.
\end{proper}

In Eq.~\eqref{eq:gentle-measurement}, the quantity $\varepsilon$ measures how far the outcome distribution is from being deterministic, while $\eta(\varepsilon)$ bounds the disturbance incurred by the gentle implementation of the measurement. In classical theory, the input can be read without changing it, so $\eta_{\rm C}=0$. In quantum theory, the instrument $\map G_a(\rho)=\sqrt{E_a}\rho\sqrt{E_a}$ associated with a POVM $(E_a)$ gives
\begin{equation}
    \label{eq:eta-quantum}
    \eta_{\rm Q}(t)=\min\left\{1,\frac t 2+\sqrt t\right\} \, .
\end{equation}
The bound $\eta_{\rm Q}$ applies to POVMs with any finite number of outcomes and follows from the gentle-measurement lemma. For binary projective measurements, we prove that the sharper bound
\begin{equation}
    \label{eq:eta-binary}
    \eta_{\rm bin}(t)=\sqrt{t(1-t)}
\end{equation}
holds for $0\leq t\leq 1/2$. This bound is useful because any process-matrix realization can, without loss of generality, be obtained using labeled projective instruments \cite{liu2025tsirelson}. These quantum disturbance bounds are derived in Methods.

To turn a state-disturbance bound into a correlation bound in a process-matrix experiment, we use the supermap realization theorem of Ref.~\cite{chiribella2008transforming}, applied after fixing the operations in all but one laboratory.
\begin{proper}[Circuit realization]
    \label{theo:superchannel}
    Every higher-order process with a single input slot that maps arbitrary input channels to channels can be realized as a one-slot circuit.
\end{proper}

In causal probabilistic theories, Property~\ref{theo:superchannel} can be derived from the stronger property OWSD. For a higher-order process with a single slot from $A$ to $B$, the requirement that inserting any channel yields a valid channel imposes the no-influence condition $B\not\to A$ on its channel representation. OWSD then decomposes this channel into a one-way-signaling circuit. This circuit realization and contractivity of distinguishability allow Eq.~\eqref{eq:gentle-measurement} to bound the change in the statistics of the process-matrix experiment caused by gentle measurements, as stated in the following lemma.

\begin{lemma}
    \label{theo:gentlesequential}
    Consider a process-matrix correlation $\pPM$ as in Eq.~\eqref{eq:general}. Fix a party $j$ and a setting $s$, and let $(\map G^{(j,s)}_a)$ be a gentle implementation of the measurement associated with the instrument $(\map M^{(j)}_{m_j\mid s})$.
    Replace the local operation by
    $
        \map M^{(j)}_{m_j\mid z_j}\circ\map G^{(j,s)}_a
    $
    and denote the resulting correlation by \mbox{$\pI(a,\bm m\mid\bm z)$}. Writing $\bm z_{\backslash j}=(z_i)_{i\ne j}$, one has the following constraints on the marginals of $\pI$:
    \begin{equation}
        \label{eq:insertion-marginal}
        \pI(a\mid\bm z)=\pPM_j(a\mid s,\bm z_{\backslash j})\, ,
    \end{equation}
    \begin{equation}
        \label{eq:insertion-distance}
        \frac12\sum_{\bm m}|\pI(\bm m\mid\bm z)-\pPM(\bm m\mid\bm z)|
        \leq\eta\big(\varepsilon_{j,s}(\bm z_{\backslash j})\big)\, ,
    \end{equation}
    where \mbox{$\varepsilon_{j,s}(\bm z_{\backslash j})
        =1-\max_a \pPM_j(a\mid s,\bm z_{\backslash j})$}.
\end{lemma}

\begin{figure}[htbp]
    \centering
    \includegraphics[width=\columnwidth]{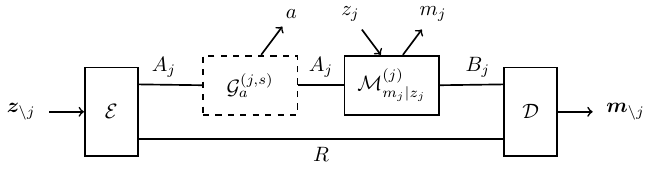}
    \caption{\emph{A gentle measurement in a local laboratory.} The higher-order process on party $j$'s operation is represented by channels $\mathcal E$ and $\mathcal D$, connected by a memory system $R$. In the local slot, the gentle instrument $\mathcal G^{(j,s)}_a$ (dashed) records the earlier outcome $a$ and returns $A_j$ with controlled disturbance. The original instrument $\mathcal M^{(j)}_{m_j\mid z_j}$ is then applied to the returned system.}
    \label{fig:gentle-insertion}
\end{figure}

Lemma~\ref{theo:gentlesequential} is proved in Methods, and the insertion of the gentle instrument is illustrated in Fig.~\ref{fig:gentle-insertion}. The gentle instrument records an earlier outcome $a$ whose marginal reproduces that of the original setting-$s$ instrument and is independent of the later choice $z_j$. It returns the input system with controlled disturbance, after which the original setting-dependent instrument is applied. The lemma can be iterated, with each disturbance evaluated in the experiment containing the preceding insertions.

To quantify the mechanism illustrated by GYNI, consider a two-party process-matrix experiment with independent, uniformly random binary settings $z_1,z_2$ and binary outcomes $m_1,m_2$. Define the marginal guessing probabilities
\begin{align}
    G_1&:=\Pr_{\pPM}(m_1=z_2\mid z_1=1)\, ,\nonumber\\
    G_2&:=\Pr_{\pPM}(m_2=z_1\mid z_2=1)\, .
    \label{eq:marginal-guessing-probabilities}
\end{align}
At the common setting $z_1=z_2=1$, define each party's error probability and outcome unpredictability, respectively, by
\begin{equation}
    \label{eq:trigger-unpredictability}
    \begin{aligned}
        \varepsilon_i&:=1-\pPM_i(1\mid1,1)\, ,\\
        \widehat\varepsilon_i&:=\min\{\varepsilon_i,1-\varepsilon_i\}\, ,
        \qquad i\in\{1,2\}\, .
    \end{aligned}
\end{equation}
Thus $\widehat\varepsilon_i$ is the probability of the less likely outcome and lies in $[0,1/2]$.

Inserting in each laboratory $i$ a gentle implementation $(\map G^{(i,1)}_{\alpha_i})$ of the setting-1 instrument before the original instrument $(\map M^{(i)}_{m_i\mid z_i})$, we obtain locally sequential statistics \mbox{$\pseq(\alpha_1,\alpha_2,m_1,m_2 \mid z_1,z_2)$}, whose marginal on $(\alpha_1,\alpha_2)$ is an LCC-compatible table distribution. Denote this marginal by $q(\alpha_1,\alpha_2\mid z_1,z_2)$ and define
\[
    \widetilde G_1:=\Pr_q(\alpha_1=z_2),
    \qquad
    \widetilde G_2:=\Pr_q(\alpha_2=z_1)\, .
\]
Since each earlier outcome is independent of its own later setting, these probabilities can equivalently be conditioned on $z_1=1$ and $z_2=1$, respectively.

LCC bounds $(\widetilde G_1+\widetilde G_2)/2$ by $3/4$. Indeed, parity erasure implies $\Pr_q(\alpha_1\oplus\alpha_2\ne z_1\oplus z_2)=1/2$. Since an incorrect parity requires at least one incorrect guess, the union bound gives $(1-\widetilde G_1)+(1-\widetilde G_2)\geq1/2$, proving the bound. Lemma~\ref{theo:gentlesequential} gives
\begin{align}
    |\widetilde G_1-G_1|&\leq\eta(\widehat\varepsilon_2)\, ,
    \label{eq:marginal-guessing-disturbance}\\
    |\widetilde G_2-G_2|&\leq\eta(\widehat\varepsilon_1)\, .
    \label{eq:marginal-guessing-disturbance-reverse}
\end{align}

Since the success probability when the other party's setting is $0$ is at most one, Eq.~\eqref{eq:marginal-guessing-probabilities} gives
\[
    G_i\leq\frac{1+\pPM_i(1\mid1,1)}{2}
    =1-\frac{\varepsilon_i}{2}\, .
\]
Combining these bounds gives the following constraint.

\begin{theo}
    \label{theo:marginal-guessing}
    The marginal guessing probabilities $G_1$ and $G_2$ defined in Eq.~\eqref{eq:marginal-guessing-probabilities} satisfy
    \begin{equation}
        \label{eq:marginal-guessing-bound}
        \frac {G_1+G_2} 2
        \leq
        \min\left\{
        \frac 3 4 + \frac{\eta(\widehat\varepsilon_1)+\eta(\widehat\varepsilon_2)}{2} \, , \, 1 - \frac{\varepsilon_1+\varepsilon_2}{4} \right\} \, .
    \end{equation}
\end{theo}

In classical theory, setting $\eta_{\rm C}=0$ in Eq.~\eqref{eq:marginal-guessing-bound} gives the tight bound $(G_1+G_2)/2\leq3/4$. Alice and Bob, the two parties in the GYNI scenario, can attain this bound using a strategy with a definite causal order: Alice guesses Bob's setting randomly and sends her own setting to Bob, who guesses it perfectly. 

In quantum theory, substituting the binary disturbance bound $\eta_{\rm bin}$ from Eq.~\eqref{eq:eta-binary} into Eq.~\eqref{eq:marginal-guessing-bound} and optimizing over $\varepsilon_1,\varepsilon_2$ yields the nontrivial bound
\begin{equation}
    \label{eq:marginal-guessing-quantum}
    \frac{G_1+G_2}{2}
    \leq\frac34+\frac{1}{2\sqrt5}
    \approx0.9736\, .
\end{equation}
To assess the tightness of this bound, we compare $(G_1+G_2)/2$ with the success probability of the lazy-guess-your-neighbor's-input (LGYNI) game. In this game, party $i$ is required to guess the other party's setting only when $z_i=1$. For independent, uniformly random binary settings, its success probability $P_{\rm LGYNI}$ is defined by
\[
\begin{aligned}
    4P_{\rm LGYNI}:={}&1+\pPM_1(0\mid1,0)\\
    &+\pPM_2(0\mid0,1)+\pPM(1,1\mid1,1)\, .
\end{aligned}
\]
The bound in Eq.~\eqref{eq:marginal-guessing-quantum} is not tight: for any fixed correlation, $(G_1+G_2)/2$ is at most $P_{\rm LGYNI}$, whose maximal value attainable by process matrices is approximately $0.8194$ \cite{liu2025tsirelson}.

Nevertheless, Eq.~\eqref{eq:marginal-guessing-quantum} constrains marginal probabilities entering the GYNI and LGYNI success probabilities and therefore implies upper bounds strictly below one for both games. Writing $P_{\rm GYNI}$ for the GYNI success probability, we have
\begin{align}
    P_{\rm GYNI}&\leq P_{\rm LGYNI}
    \leq\frac12+\frac{G_1+G_2}{4} \nonumber \\
    &\leq\frac78+\frac{1}{4\sqrt5}\approx0.9868\, .
    \label{eq:gyniscore}
\end{align}
Detailed derivations of the bounds \eqref{eq:marginal-guessing-quantum} and \eqref{eq:gyniscore}, together with the proof of Theorem~\ref{theo:marginal-guessing}, are given in Methods.

We apply the insertion of gentle instruments iteratively to obtain constraints on process-matrix correlations in arbitrary finite scenarios. Starting from a realization of a correlation $\pPM$ close to a deterministic target $\pD$, we sequentially insert gentle instruments for every local setting. The resulting tables satisfy LCC, and a local readout yields a classical correlation $\pC$ whose distance from $\pD$ is controlled by the accumulated disturbance. 
We measure the distance between two correlations by the worst-case total variation distance over settings,
\[
    d_\infty(p,q):=\max_{\bm z}\frac12\sum_{\bm m}
    |p(\bm m\mid\bm z)-q(\bm m\mid\bm z)|\, .
\]
The function $\gamma$ in the following theorem bounds the classical approximation error in terms of the original error:
\[
    d_\infty(\pC,\pD) \leq \gamma\left(d_\infty(\pPM,\pD)\right) \, ,
\]
and inverting it yields a lower bound on the original error.

\begin{theo}
    \label{theo:generalnontrivial}
    Let $\pPM$ be the statistics of a process-matrix experiment and let $\pD$ be a deterministic correlation in the same finite scenario. Then
\begin{equation}
    \label{eq:gap}
    d_\infty(\pPM,\pD)\geq \gamma^{-1}\left(\min_{\pC} d_\infty(\pC,\pD)\right)\, ,
\end{equation}
where $\pC$ ranges over the statistics of classical process-matrix experiments and $\gamma$ is defined as follows, with $\eta$ the disturbance bound of Property~\ref{prop:gentle-measurement}:
\[
\begin{aligned}
K&=\sum_{i=1}^{N}|\mathcal Z_i|,\qquad t_0=\varepsilon\, ,\\
t_{k+1}&=t_k+\eta\big(\min\{t_k,1\}\big),\quad 0\leq k<K\, ,\\
\gamma(\varepsilon)&=N t_K\, .
\end{aligned}
\]
\end{theo}

For a fixed finite scenario, the LCC-compatible table distributions form a compact polytope whose image under a local readout is exactly the classical correlation set. Thus, if $\pD$ has no classical realization, $\min_{\pC}d_\infty(\pC,\pD)>0$. Since $\gamma$ is continuous and strictly increasing with $\gamma(0)=0$, the right-hand side of Eq.~\eqref{eq:gap} is then strictly positive. The theorem therefore excludes an open neighborhood of $\pD$ from the quantum process-matrix correlation set, strengthening exact nonrealizability to a finite separation. The proof, including the recursive disturbance estimate, is given in Methods.

Theorems~\ref{theo:marginal-guessing} and \ref{theo:generalnontrivial} extend directly to general causal probabilistic theories satisfying the gentle-measurement and circuit-realization properties (Properties~\ref{prop:gentle-measurement} and \ref{theo:superchannel}), with the trace norm replaced by the operational norm \cite{chiribella2010probabilistic}. If Property~\ref{theo:superchannel} holds only for a restricted class of higher-order processes, the conclusions apply to that class.

\section{Discussion}

Our results establish LCC as an information-theoretic foundation for the process matrix formalism, linking causal consistency within laboratories to the structure and observable correlations of higher-order processes. Consistency under arbitrary local feed-forward is equivalent to parity-erasure conditions that, through local process tomography, characterize the channels representing valid process matrices. LCC also gives an exact characterization of classical process-matrix correlations through local readout of outcome-table distributions. For quantum correlations, LCC and gentle measurements yield bounds on two-way guessing success and exclude neighborhoods of deterministic correlations without classical realizations.

These results suggest further investigation of the implications of LCC and other information-theoretic principles for process-matrix correlations. A first question is whether LCC alone yields nontrivial constraints on general process-matrix correlations. Another is whether the gentle-measurement property can give tighter constraints than those derived here. Additional information-theoretic principles, combined with LCC, may further restrict causal-inequality violations.

LCC may also clarify the operational basis of causal modeling and computation with indefinite causal structure. In cyclic quantum causal models \cite{barrett2021cyclic,ferradini2025cyclic}, one aim is to derive graph-theoretic admissibility conditions directly from LCC. An example is the ``siblings-on-cycles'' condition: every directed cycle in a faithful, consistent quantum causal model contains two vertices with a common parent \cite{tselentis2023admissible}. For computational models with indefinite causal structure \cite{araujo2017quantumcomputation,abbott2024quantum}, a related question is whether and how LCC translates into complexity bounds and thereby restricts the computational advantages of such models.

Finally, the characterization could be adapted to more structured process classes and broader operational frameworks. For quantum circuits with classical or quantum control of causal order \cite{wechs2021quantum}, additional operational constraints could identify the corresponding subclasses within the process-matrix set. Multi-round process matrices \cite{hoffreumon2021multi} may require a hierarchy of parity-erasure conditions reflecting the successive stages of local operations. In bistochastic quantum theory \cite{chiribella2022quantum}, an analogue of LCC might clarify an information-theoretic symmetry of input-output direction.

\section{Methods}

\subsection*{Operational framework and notation}
It is straightforward to generalize the process matrix formalism to any operational probabilistic theory (OPT) \cite{oreshkov2016causal}. In a general OPT \cite{hardy2001quantum,barrett2007information,barnum2007generalized,chiribella2010probabilistic,chiribella2011informational,hardy2011foliable,hardy2013formalism,chiribella2015quantum,hardy2016quantum,d2017quantum}, the state space of a system $A$ is usually assumed to be a convex set $\St(A)$ which spans a finite-dimensional vector space $\St_\R(A)$. Every state in $\set{St}(A)$ can be determined by the statistics of a choice of effects that span the dual space of $\St_\R(A)$. 
Channels are defined as deterministic transformations between systems, while a test corresponds to a refinement of a channel (analogous to an instrument in quantum theory). In a causal OPT, $u^A$ denotes the unique deterministic effect on $A$, and a deterministic state $\rho$ satisfies $u^A(\rho)=1$. We refer to Refs. \cite{chiribella2010probabilistic,chiribella2015quantum,d2017quantum} for further details on the OPT framework.
The general process matrix formalism describes supermaps that are completely positive transformations of channels. That is, given a supermap $\map S$, for any choice of channels $\widetilde{\map M}^{(i)}$ from the composite system $A_i A'_i$ to the composite system $B_i B'_i$, the map $\map S \otimes \map I^{\Transf(A' \to B')}$ produces a valid channel from system $A' := A'_1\ldots A'_N$ to system $B' := B'_1\ldots B'_N$, where $\map I^{\Transf(A' \to B')}$ denotes the identity map on the set of transformations from $A'$ to $B'$. We adopt the physicalization of readout \cite{chiribella2014dilation}, that is, every test can be realized as a channel followed by a measurement on an auxiliary system, so that supermaps can be defined by their action on channels.

\subsection*{Parity erasure and classical process validity}
An $N$-party process matrix $S$ acting on system $A_1 \ldots A_N B_1 \ldots B_N$ can be characterized by the following conditions \cite{araujo2015witnessing}:
\begin{align}
    &S \geq 0 \, , \nonumber\\
    &\Tr(S) = \prod_{i=1}^N d_{B_i} \, , \nonumber\\
    &\forall ~\text{nonempty}~ I \subseteq \{ 1,\ldots, N \} \nonumber \\
    & {}_{[\prod_{i \in I}(1-B_i) \prod_{i\notin I} A_i B_i ]} S = 0 \, , \label{eq:PM3}
\end{align}
where we have used the notation ${}_{[X]}S := \Tr_X[S] \otimes \frac{\id_X}{d_X}$ for a quantum system $X$ of dimension $d_X$, and the shorthand ${}_{[\sum \alpha_X X]}S := \sum_X \alpha_X \cdot {}_{[X]}S$ for a linear combination of the terms $\{ {}_{[X]}S \}_X$ associated with a collection of systems indexed by $X$ (for example, ${}_{[(1-X)Y]}S$ serves as a shorthand for ${}_{[Y]}S - {}_{[XY]}S$).

To prove necessity in Theorem~\ref{theo:parity}, fix $\bm x$ and assume LCC. The normalization requirement in Eq.~\eqref{eq:classical} is
\[
    \sum_{\bm a, \bm y} 
    \pseq(\bm a \mid \bm x, \bm y)
    \prod_{i=1}^N \kappa_i(y_i \mid a_i)
    = 1
\]
for every collection of local feedback channels $\kappa_i$. The corresponding diagonal classical process matrix is
\[
    S_{\bm x}
    :=
    \sum_{\bm a, \bm y}
    \pseq(\bm a \mid \bm x, \bm y)
    \, |\bm a, \bm y\rangle\langle \bm a, \bm y| \, .
\]
For binary later settings, applying the constraints in Eq.~\eqref{eq:PM3} to $S_{\bm x}$ yields
\begin{align}
    &\forall~\text{nonempty } I \subseteq \{1,\ldots,N\} \nonumber \\
    &\sum_{\bm a_I, \bm y}
    \prod_{i\in I} (-1)^{y_i}
    \pseq(\bm a_I \mid \bm x, \bm y)
    |\bm a_I\rangle\langle \bm a_I|
    = 0\, . \nonumber
\end{align}
Linear independence of the projectors $|\bm a_I\rangle\langle\bm a_I|$ forces each coefficient to vanish. This gives Eq.~\eqref{eq:pe} for the identity encodings $f_i(w_i)=w_i$, proving necessity for binary settings. For arbitrary setting alphabets and encoding functions $f_i$, define
\[
    q_f(\bm a\mid\bm x,\bm w)
    :=\pseq(\bm a\mid\bm x,f(\bm w))\, .
\]
Any local channel $\lambda_i(w_i\mid a_i)$ induces the channel
\[
    \kappa_i(y_i\mid a_i)
    :=\sum_{w_i:f_i(w_i)=y_i}\lambda_i(w_i\mid a_i)\, .
\]
Thus LCC for $\pseq$ implies LCC for $q_f$. Applying the binary result to $q_f$ gives Eq.~\eqref{eq:pe} for every collection of encodings.

For the if direction, fix $\bm x$ and interpret $\pseq(\bm a\mid\bm x,\bm y)$ as the LPT representation of a classical channel, obtained through basis-state preparations and readout. Classical probability theory is causal and satisfies local tomography, OWSD, and physicalization of readout. The implication $(3)\implies(1)$ of Theorem~\ref{theo:characterization}, whose proof below does not rely on this direction, therefore applies. The parity-erasure conditions ensure that the channel represents a valid classical process matrix, so Eq.~\eqref{eq:classical} is normalized for every collection of local classical channels. Nonnegativity then completes the proof.

\subsection*{Recursive characterization of process matrices}
We work in the finite-dimensional causal OPT framework above, with local tomography and OWSD as assumed in the main text.

To prove Theorem~\ref{theo:characterization}, it suffices to establish $(3)\implies(1)$: the implications $(1)\implies(2)\implies(3)$ follow from the definitions and the necessity direction of Theorem~\ref{theo:parity}. We use the following definition, which includes the auxiliary systems needed for the recursion:
\begin{defi}
    A channel from system $B_1 \ldots B_k P$ to system $A_1 \ldots A_k F$ is $k$-parity-erasing if, for any fixed input on $P$ and after tracing out $F$, the LPT representation of the reduced $k$-partite channel satisfies the parity-erasure conditions \eqref{eq:pe}.
\end{defi}
We insert one local channel at a time into a parity-erasing channel. By local tomography, singleton parity erasure implies $B_i\not\to A_i$, so OWSD supplies a circuit with a slot from $A_i$ to $B_i$. The following lemma shows that filling this slot preserves the remaining parity-erasure conditions:
\begin{lemma}
    \label{theo:induction}
    Let $\map T$ be a $k$-parity-erasing channel from system $B_1\ldots B_kP$ to system $A_1\ldots A_kF$. Then, for any channel $\map C$ from $A_kP'$ to $B_kF'$, inserting $\map C$ into the one-way-signaling circuit of $\map T$ yields a $(k-1)$-parity-erasing channel from $B_1\ldots B_{k-1}PP'$ to $A_1\ldots A_{k-1}FF'$.
\end{lemma}
Starting with trivial $P,F$, apply the lemma successively to arbitrary extended local channels from $A_iA_i'$ to $B_iB_i'$. After all slots are filled, the result is a physical channel from $A_1'\ldots A_N'$ to $B_1'\ldots B_N'$. Physicalization of readout then gives valid statistics for arbitrary local tests. This proves $(3)\implies(1)$ and completes the characterization. Applying local tomography to the reduced outputs converts the probability constraints into the state equalities of Corollary~\ref{theo:rac}.

\subsection*{Proof of Lemma~\ref{theo:induction}}
\label{app:2}

We use the linear extension of transformations from $B$ to $A$ to maps from $\St_\mathbb{R}(B)$ to $\St_\mathbb{R}(A)$, and the deterministic effects $u^A$ introduced above. For brevity, write
\[
    u^{\overline I} := \bigotimes_{i \notin I} u^{A_i},
    \qquad
    \rho^{\overline I} := \bigotimes_{i \notin I} \rho^{B_i}_i\, .
\]

We begin with an equivalent definition of parity-erasing channels.

\begin{prop}
\label{theo:parityerasing}
A channel $\map T$ from system $B_1\ldots B_k P$ to system $A_1\ldots A_k F$ is $k$-parity-erasing if and only if, for every nonempty subset $I \subseteq \{1,\ldots,k\}$ and for arbitrary deterministic states on the corresponding systems,
\begin{equation}
\label{eq:b1}
u^{\overline I}u^{F}\,
\map T
\bigg(
\bigotimes_{i\in I}\big(\rho^{B_i}_{i,0}-\rho^{B_i}_{i,1}\big)
\otimes
\rho^{\overline I}
\otimes
\sigma^P
\bigg)
=0\, .
\end{equation}
\end{prop}

\Proof
Let $\big\{\rho^{B_i}_{y_i} \big\}$ be an informationally complete set of deterministic states on system $B_i$, i.e., a set spanning the vector space $\St_\mathbb{R}(B_i)$.

Applying Eq.~\eqref{eq:pe} to these local probes, and using local tomography and linearity, gives the equivalent condition below for every nonempty $I\subseteq\{1,\ldots,k\}$, every collection of encodings $f_i$, and every deterministic input $\sigma^P$:
\begin{equation}
\label{eq:b2}
u^{\overline I}u^{F}\,
\map T
\bigg(
\bigotimes_{i=1}^k P_{f_i}^{B_i}
\otimes
\sigma^P
\bigg)
=0\, ,
\end{equation}
where
\[
P_{f_i}^{B_i}
:=
\begin{cases}
\rho^{B_i}_{f_i(0)}-\rho^{B_i}_{f_i(1)}, & i\in I\, , \\[6pt]
\rho^{B_i}_{f_i(0)}+\rho^{B_i}_{f_i(1)}, & i\notin I\, .
\end{cases}
\]

For $i\notin I$, the vectors $\big\{P_{f_i}^{B_i}\big\}_{f_i}$ span the full space $\St_\mathbb{R}(B_i)$.  
For $i\in I$, the span satisfies
\[
\Span\big\{P_{f_i}^{B_i}\big\}_{f_i}
=
\{\,Q\in \St_\mathbb{R}(B_i): u^{B_i}(Q)=0\,\}\, .
\]
Therefore Eq.~\eqref{eq:b2} holds for all choices of $f_i$ if and only if Eq.~\eqref{eq:b1} holds for arbitrary deterministic states on the corresponding systems. \qed

In the remainder of this proof, for $I\subseteq\{1,\ldots,k-1\}$ set $J:=\{1,\ldots,k-1\}\setminus I$, and write
\[
    u^{A_J}:=\bigotimes_{j\in J}u^{A_j},\qquad
    \rho^{B_J}:=\bigotimes_{j\in J}\rho_j^{B_j}\, .
\]
The removed laboratory $k$ is therefore excluded from these products.

Let $\map T_{\map C}$ denote the channel obtained by inserting $\map C$ into the one-way-signaling circuit of $\map T$. It suffices to show that, for any nonempty subset $I\subseteq\{1,\ldots,k-1\}$ and for arbitrary deterministic states on the corresponding systems,
\begin{equation}
\label{eq:zero}
u^{A_J}u^{F}u^{F'}\,
\map T_{\map C}
\bigg(
\bigotimes_{i\in I}\big(\rho^{B_i}_{i,0}-\rho^{B_i}_{i,1}\big)
\otimes
\rho^{B_J}
\otimes
\sigma^P
\otimes
\xi^{P'}
\bigg)
=0\, .
\end{equation}

It is enough to establish Eq.~\eqref{eq:zero} for product ancillary inputs $\sigma^P\otimes\xi^{P'}$: local tomography implies that these products span $\St_\mathbb{R}(PP')$, so linearity extends the identity to arbitrary joint states on $PP'$.

Let $\map T_\sigma$ denote the channel obtained from $\map T$ by preparing $\sigma$ on system $P$ and discarding the output system $F$. Similarly, let $\map C_\xi$ denote the channel obtained from $\map C$ by preparing $\xi$ on system $P'$ and discarding the output system $F'$.
Since local tomography allows for a local characterization of transformations \cite{chiribella2010probabilistic}, $\map C_\xi$ can be written as a linear combination of parallel compositions of an effect on $A_k$ and a state on $B_k$:
\[
\map C_\xi(\cdot)
=
\sum_j
c_j\, e_j^{A_k}(\cdot)\,\rho^{B_k}_{k,j} \, ,
\]
where $(c_j)$ are real coefficients and $\big(\rho^{B_k}_{k,j} \big)$ are deterministic states.
Introducing an arbitrary deterministic state $\pi^{B_k}$, we can rewrite this decomposition as
\[
\map C_\xi(\cdot)
=
u^{A_k}(\cdot)\,\pi^{B_k}
+
\sum_j
c_j\, e_j^{A_k}(\cdot)\big(\rho^{B_k}_{k,j}-\pi^{B_k}\big)\, .
\]
Here we have used the following property of causality:
\[
u^{A_k} = u^{B_k} \circ \map C_\xi = \sum_j c_j\, e_j^{A_k}\, .
\]

Therefore, to prove Eq.~\eqref{eq:zero}, it suffices to verify the following two conditions:
\begin{equation}
\label{eq:condition1}
u^{A_J}u^{A_k}\,
\map T_\sigma
\bigg(
\bigotimes_{i\in I}
(\rho^{B_i}_{i,0}-\rho^{B_i}_{i,1})
\otimes
\rho^{B_J}
\otimes
\pi^{B_k}
\bigg)
=0\, ,
\end{equation}
and
\begin{equation}
\label{eq:condition2}
u^{A_J}\,
\map T_\sigma
\bigg(
\bigotimes_{i\in I}
(\rho^{B_i}_{i,0}-\rho^{B_i}_{i,1})
\otimes
\rho^{B_J}
\otimes
(\rho^{B_k}_{k,j}-\pi^{B_k})
\bigg)
=0\, .
\end{equation}

By Proposition~\ref{theo:parityerasing}, Eq.~\eqref{eq:condition1} is the parity-erasure condition for $I$, while Eq.~\eqref{eq:condition2} is that for $I\cup\{k\}$. Applying $e_j^{A_k}$ to the latter and using the decomposition of $\map C_\xi$ proves Eq.~\eqref{eq:zero}, completing the proof of Lemma~\ref{theo:induction}.

\subsection*{Gentle implementation of quantum measurements}
\label{methods:gentle-quantum}
We derive the quantum disturbance bounds for the L\"uders instrument specified in the main text. 
Choose a most likely outcome $a_*$, write $E=E_{a_*}$ and $t=1-\Tr(E\rho)$. The gentle-measurement lemma states that
\[
\frac 1 2 \left\| \rho - \sqrt E \rho \sqrt E \right\|_1 \leq \sqrt t \, .
\]
The remaining term $\sum_{a \neq a_*} \sqrt{E_a} \rho \sqrt{E_a}$ forms a positive operator of trace $t$.
The triangle inequality bounds the non-selective disturbance by $\sqrt t+t/2$. Both the original and non-selective output states are normalized, so their trace distance is also at most one. Combining these estimates yields $\eta_{\rm Q}$ in Eq.~\eqref{eq:eta-quantum}.

For a binary projective measurement $(P,Q)$, where $Q=\id-P$, dephasing removes only the two off-diagonal blocks of $\rho$. With $P$ denoting the more likely outcome and $t=\Tr(Q\rho)$,
\begin{align*}
    \frac12\|P\rho P+Q\rho Q-\rho\|_1
    &=\|P\rho Q\|_1\\
    &\leq\|P\sqrt\rho\|_2\|\sqrt\rho Q\|_2\\
    &=\sqrt{t(1-t)}\, .
\end{align*}
The inequality is H\"older's inequality for Schatten norms, $\|XY\|_1\leq\|X\|_2\|Y\|_2$, while the final equality follows from $\|P\sqrt\rho\|_2^2=\Tr(P\rho)=1-t$ and $\|\sqrt\rho Q\|_2^2=\Tr(Q\rho)=t$. Since $P$ denotes the more likely outcome, $0\leq t\leq1/2$. 

The bound applies within a larger experiment by treating the measured component and any untouched components as a single composite system. In quantum theory, the measurement operators are then $E_a\otimes\id$, whose L\"uders maps are precisely $\map G_a\otimes\map I$, since $\sqrt{E_a\otimes\id}=\sqrt{E_a}\otimes\id$. The outcome probabilities, and hence $\varepsilon$, depend only on the state of the measured component. Thus the same bound follows directly from the single-system statement. This establishes the reference-system statement of Property~\ref{prop:gentle-measurement}.

\subsection*{Gentle insertion bounds}
To prove Lemma~\ref{theo:gentlesequential}, fix $\bm z_{\backslash j}$ and use the one-slot circuit from Property~\ref{theo:superchannel}, illustrated in Fig.~\ref{fig:gentle-insertion}. Let $\omega^{A_jR}_{\bm z_{\backslash j}}$ be the state on the slot input and circuit memory; it is independent of $z_j$.

The measurement associated with the setting-$s$ instrument in laboratory $j$ is the collection of effects
\[
    e^{(j,s)}_a:=u^{B_j}\circ\map M^{(j)}_{a\mid s}\, ,
\]
where $u^{B_j}$ is the deterministic effect on the output system. The gentle implementation satisfies $u^{A_j}\circ\map G^{(j,s)}_a=e^{(j,s)}_a$. Consequently, after all later outcomes are discarded, the probability of the inserted outcome is
\begin{align*}
    \pI(a\mid\bm z)
    &=\big(e^{(j,s)}_a\otimes u^R\big)
      \big(\omega^{A_jR}_{\bm z_{\backslash j}}\big)\\
    &=\pPM_j(a\mid s,\bm z_{\backslash j})\, .
\end{align*}
This proves Eq.~\eqref{eq:insertion-marginal}.

If the inserted outcome is discarded, the state passed to the remainder of the circuit changes from $\omega^{A_jR}_{\bm z_{\backslash j}}$ to
\[
    \omega'^{A_jR}_{\bm z_{\backslash j}}
    :=\left[\left(\sum_a\map G^{(j,s)}_a\right)
    \otimes\map I^R\right]
    \left(\omega^{A_jR}_{\bm z_{\backslash j}}\right)\, .
\]
Let $\Eff(A)$ denote the set of effects on system $A$. For $\delta\in\St_\mathbb{R}(A)$, we use the operational norm convention \cite{chiribella2010probabilistic}
\[
    \|\delta\|_{\rm op}
    :=\sup_{e\in\Eff(A)}e(\delta)-\inf_{e\in\Eff(A)}e(\delta)\, ,
\]
which reduces to the trace norm in quantum theory. For differences of normalized states, $u^A(\delta)=0$ and complementary effects give $\tfrac12\|\delta\|_{\rm op}=\sup_{e\in\Eff(A)}|e(\delta)|$. Thus the operational distance is half the norm, consistently with total variation distance for classical outcomes.

The gentle-measurement property, in its operational-probabilistic form, gives
\begin{equation}
    \label{eq:insertion-operational-distance}
    \frac12\left\|
    \omega'^{A_jR}_{\bm z_{\backslash j}}
    -\omega^{A_jR}_{\bm z_{\backslash j}}
    \right\|_{\rm op}
    \leq\eta\big(\varepsilon_{j,s}(\bm z_{\backslash j})\big)\, .
\end{equation}
The subsequent circuit is a test with classical outcome $\bm m$, so the total variation distance between the two outcome distributions cannot exceed the operational distance in Eq.~\eqref{eq:insertion-operational-distance}. This proves Eq.~\eqref{eq:insertion-distance}.

Existing setting-independent insertions can be absorbed into the circuit, with their records retained in $R$ or among the final outcomes. The same argument then bounds the joint distribution of these records and the original outcomes, with $\varepsilon_{j,s}$ evaluated in the experiment containing those insertions. This justifies the iterative use of the lemma in the main text.

\subsection*{Marginal guessing and game bounds}

By Lemma~\ref{theo:gentlesequential}, inserting the gentle instrument in laboratory $2$ changes the laboratory-$1$ outcome distribution at $z_1=1$ by at most $\eta(\widehat\varepsilon_2)$, where $\widehat\varepsilon_2$ is the unpredictability at the common setting $z_1=z_2=1$. The insertion in laboratory $1$ then records the current setting-$1$ probabilities exactly, proving Eq.~\eqref{eq:marginal-guessing-disturbance}. Exchanging the laboratory labels gives Eq.~\eqref{eq:marginal-guessing-disturbance-reverse}; both estimates concern the same doubly inserted experiment. Together with the parity-erasure and elementary guessing bounds established in the main text, they prove Theorem~\ref{theo:marginal-guessing}.

To obtain the quantum bound in Eq.~\eqref{eq:marginal-guessing-quantum}, substitute $\eta_{\rm bin}$ from Eq.~\eqref{eq:eta-binary} into Eq.~\eqref{eq:marginal-guessing-bound}. Since $\widehat\varepsilon_i\in[0,1/2]$ is either $\varepsilon_i$ or $1-\varepsilon_i$, we have $\eta_{\rm bin}(\widehat\varepsilon_i)=\sqrt{\varepsilon_i(1-\varepsilon_i)}$. Setting $u=(\varepsilon_1+\varepsilon_2)/2\in[0,1]$ and using the concavity of the function $\sqrt{t(1-t)}$, together with the elementary bounds on $G_i$ in the main text, gives
\[
    G_1+G_2
    \leq\min\left\{\frac32+2\sqrt{u(1-u)},\,2-u\right\}\, .
\]

For $u\geq1/2$, the bound $2-u$ is at most $3/2$. On $0\leq u\leq1/2$, $2-u$ decreases while $3/2+2\sqrt{u(1-u)}$ increases. Their intersection occurs at $u=1/2-1/\sqrt5$ and yields $G_1+G_2\leq3/2+1/\sqrt5$, proving Eq.~\eqref{eq:marginal-guessing-quantum}.

For the game-score comparisons, use the definition of $P_{\rm LGYNI}$ given in the main text. At the common trigger setting, the union bound gives
\[
\begin{aligned}
    &\pPM_1(1\mid1,1)+\pPM_2(1\mid1,1)\\
    &\qquad\leq1+\pPM(1,1\mid1,1)\, .
\end{aligned}
\]
Together with the definitions of $G_1,G_2$, this implies $(G_1+G_2)/2\leq P_{\rm LGYNI}$, establishing the comparison used in the main text.

For the upper bound on $P_{\rm LGYNI}$, use $\pPM(1,1\mid1,1)\leq[\pPM_1(1\mid1,1)+\pPM_2(1\mid1,1)]/2$ and $\pPM_1(0\mid1,0)+\pPM_2(0\mid0,1)\leq2$ in the score formula given in the main text. This gives $4P_{\rm LGYNI}\leq2+G_1+G_2$. Since winning GYNI also wins LGYNI, Eq.~\eqref{eq:marginal-guessing-quantum} gives the game bounds in Eq.~\eqref{eq:gyniscore}.

\subsection*{Classical approximation of nearly deterministic correlations}
To prove Theorem~\ref{theo:generalnontrivial}, write $\pD(\bm m\mid\bm z)=\delta_{\bm m,g(\bm z)}$ and set $\varepsilon=d_\infty(\pPM,\pD)$. We use $K$, $t_k$, and $\gamma$ as defined in the theorem. Each update $t\mapsto t+\eta(\min\{t,1\})$ is continuous and strictly increasing, with value zero at zero. Thus $\gamma$ is continuous and strictly increasing, with $\gamma(0)=0$. Since $t_K\geq t_0$, we also have $\gamma(1)\geq N\geq1$, ensuring that $\gamma^{-1}$ is defined on $[0,1]$.

Starting from a realization of $\pPM$, enumerate all pairs $(i,s)$, with $s\in\mathcal Z_i$, in any fixed order. At step $k+1$, insert a gentle implementation of $e^{(i,s)}_a=u\circ\map M^{(i)}_{a\mid s}$ after all earlier insertions in laboratory $i$ and before its original setting-dependent instrument. Record its outcome as $\alpha_{i,s}$. All inserted instruments have fixed settings and form a prefix that is independent of the later choice $z_i$.

Let $p^{(k)}$ denote the joint distribution of the original outcomes and the first $k$ records. We prove inductively, for every $\bm z$, that
\begin{align}
    \Pr_{p^{(k)}}[\bm m\ne g(\bm z)\mid\bm z]&\leq t_k\, ,
    \label{eq:induction-original}\\
    \Pr_{p^{(k)}}[\alpha_{i,s}\ne g_i(s,\bm z_{\backslash i})\mid\bm z]&\leq t_k\, ,
    \label{eq:induction-record}
\end{align}
for every record already inserted. At $k=0$, Eq.~\eqref{eq:induction-original} holds because
$t_0=\varepsilon=d_\infty(\pPM,\pD)=\max_{\bm z}\Pr_{\pPM}[\bm m\ne g(\bm z)\mid\bm z]$,
and Eq.~\eqref{eq:induction-record} is vacuous.

Suppose the statements hold at step $k$. At the next insertion $(i,s)$, the probability of the new outcome equals the current local marginal at the setting $z_i=s$, by the prefix version of Eq.~\eqref{eq:insertion-marginal}. Since $\{m_i\ne g_i(s,\bm z_{\backslash i})\}\subseteq\{\bm m\ne g(s,\bm z_{\backslash i})\}$, Eq.~\eqref{eq:induction-original} at the setting $z_i=s$ implies that the new record has error at most $t_k$ relative to $g_i(s,\bm z_{\backslash i})$. Its unpredictability is therefore at most $t_k$. The non-selective insertion changes the joint distribution of all old records and original outcomes by at most $\eta(\min\{t_k,1\})$, by the prefix version of Eq.~\eqref{eq:insertion-distance} and the monotonicity of $\eta$. Each old error probability consequently increases by at most this amount, while the new record's error is at most $t_k$; hence both statements hold at step $k+1$ by the recursion in Theorem~\ref{theo:generalnontrivial}. When $t_k\geq1$, both statements are trivial, and $\min\{t_k,1\}$ keeps the argument of $\eta$ in its domain.

After $K$ insertions, discard the original outcomes and retain the table distribution \mbox{$\pT(\bm\alpha\mid\bm z)$}. The entire prefix in each laboratory is a single earlier instrument, and the original operation is a later instrument; therefore $\pT$ satisfies LCC and defines a valid classical process. Read out $\alpha_{i,z_i}$ in each laboratory, as in Eq.~\eqref{eq:classical-embedding}, to obtain a classical process-matrix correlation $\pC_*$. By the union bound applied to Eq.~\eqref{eq:induction-record},
\[
    1-\pC_*(g(\bm z)\mid\bm z)
    \leq\sum_i\Pr[\alpha_{i,z_i}\ne g_i(\bm z)\mid\bm z]
    \leq N t_K\, .
\]
For a deterministic target, the left-hand side is the total variation distance at setting $\bm z$, so $d_\infty(\pC_*,\pD)\leq Nt_K=\gamma(\varepsilon)$. Consequently,
\[
    \min_{\pC} d_\infty(\pC,\pD)
    \leq d_\infty(\pC_*,\pD)
    \leq \gamma(\varepsilon)\, .
\]
Applying the strictly increasing inverse $\gamma^{-1}$ yields Eq.~\eqref{eq:gap}.

To justify the compactness statement in the main text, observe that table distributions lie in a finite product of probability simplices, with one linear LCC constraint for each deterministic local feedback map. Their image under the linear readout in Eq.~\eqref{eq:classical-embedding} is therefore compact.

Only LCC, the gentle-measurement and circuit-realization properties, and contractivity of operational distinguishability enter this construction, which justifies the theory-independent extension stated in the main text.

\section*{Data Availability}

No datasets were generated or analyzed during the current study.

\section*{Code Availability}

No custom code was used to generate or support the results reported in this study.

\bibliography{references}

\section*{Acknowledgments}

The authors are grateful to G. Chiribella and R. Kunjwal for helpful comments and to A. Grinbaum, C. Branciard, M. Wilson, V. Vilasini, J. Bavaresco, and J. Surace for stimulating discussions. This work was supported by the F.R.S.-FNRS under project CHEQS within the Excellence of Science (EOS) program, by the ID\# 62312 grant from the John Templeton Foundation, as part of the \href{https://www.templeton.org/grant/the-quantum-information-structure-of-spacetime-qiss-second-phase}{``The Quantum Information Structure of Spacetime'' Project (QISS)}, and by the ID\# 63683 grant from the John Templeton Foundation, as part of the \href{https://www.withoutspacetime.org}{``WithOut SpaceTime'' Project (WOST)}. The opinions expressed in this publication are those of the authors and do not necessarily reflect the views of the John Templeton Foundation. O.O. is a Senior Research Associate of the Fonds de la Recherche Scientifique (F.R.S.-FNRS).

\medskip
\section*{Author Contributions}

Z.L. derived the parity-erasure conditions, related them to causality within local laboratories, and established the correlation bounds. O.O. extended the parity-erasure characterization of process matrices to general locally tomographic theories. Z.L. and O.O. wrote the manuscript.

\medskip

\section*{Competing Interests}

The authors declare no competing interests.

\end{document}